\documentclass[aps,prd,floatfix,twocolumn,nofootinbib,superscriptaddress,showkeys, preprintnumbers,10pt]{revtex4-2}

\usepackage{amsmath,amssymb,bm,color,float,graphicx,mathrsfs}
\usepackage{hyperref}
\usepackage{Macro}
\hypersetup{colorlinks=true,linkcolor=blue,citecolor=blue,urlcolor=blue}
\begin{document}

% preprint number %
\preprint{TU-1318}
\preprint{RESCEU-21/26}

% Title %
\title{Cosmic birefringence tomography with radio galaxies I: Testing birefringence solutions to the CMB--BAO tension}

\author{Toshiya Namikawa}
\affiliation{Kavli IPMU (WPI), UTIAS, The University of Tokyo, Kashiwa, 277-8583, Japan}

\author{Fumihiro Naokawa}
\affiliation{Department of Physics, Tohoku University, Sendai, Miyagi 980-8578, Japan}
\affiliation{Departamento de F\'isica de la Tierra y Astrof\'isica, Universidad Complutense de Madrid, 28040 Madrid, Spain}
\affiliation{Research Center for the Early Universe, The University of Tokyo, Bunkyo-ku, Tokyo 113-0033, Japan}

\author{Richard Battye}
\affiliation{Jodrell Bank Centre for Astrophysics, Department of Physics and Astronomy, University of Manchester, Manchester, M13 9PL, U.K}

\author{Salvatore Samuele Sirletti}
\affiliation{Kavli IPMU (WPI), UTIAS, The University of Tokyo, Kashiwa, 277-8583, Japan}

% Date %
\date{\today}

% Abstract %
\begin{abstract}
We present an independent test of cosmic birefringence induced by axionlike particles (ALPs) at low redshift using radio galaxies. 
We focus on models that can alleviate the mild tension between recent baryon acoustic oscillation (BAO) measurements and observations of the cosmic microwave background (CMB). 
This tension can be explained by a large cosmic birefringence rotation angle generated during the reionization epoch. 
Such a large rotation can arise if the recent isotropic cosmic birefringence measurements are subject to an $n\pi$ phase ambiguity, with $n\in\mathbb{Z}$, or if cold dark matter injects energy into ALPs.
We constrain these two scenarios by fitting the cosmic birefringence angles estimated from the jet morphologies and polarization angles of radio galaxies obtained by Battye, Jackson \& Browne (2026). 
For the $n\pi$ model, the radio-galaxy data exclude $|n|\geq 1$ at more than $2\sigma$ for $\log_{10}(m_{\phi}/{\rm eV})\lesssim -30.7$. 
For the energy injection model, the data are consistent with no energy injection, but with a slight relative preference for a nonzero energy injection compared to the no-injection case.
Since our samples are limited to $\sim 3000$ and there could still be several astrophysical uncertainties and statistical fluctuations, further exploration of tomographic cosmic birefringence using a large number of low-redshift radio galaxies is important for independent tests of cosmic birefringence. 
\end{abstract} 


%////////////////////////////////////////%
% MAIN MATTER 
%////////////////////////////////////////%

% Contents %
\maketitle

%////////////////////////////////////////%
\section{Introduction} \label{sec:intro}
%////////////////////////////////////////%

% recent measurements of cosmic birefringence from CMB
Recent analyses of cosmic microwave background (CMB) polarization data have revealed an intriguing hint of cosmic birefringence, a phenomenon in which the plane of linear polarization rotates as light propagates through the Universe, potentially providing a window into physics beyond the standard cosmological model~\cite{Nakai:2023} (see Refs.~\cite{Alighieri:2015:review,Komatsu:2022:review} for reviews). 
Using the method proposed by Minami et al.~\cite{Minami:2019:rot}, in which Galactic foreground emission is employed to calibrate the absolute instrumental polarization angle, Minami \& Komatsu \cite{Minami:2020:biref} first reanalyzed \textit{Planck} polarization data and reported a hint for isotropic birefringence at $2.4\,\sigma$ statistical significance. The analysis is improved by adding Wilkinson Microwave Anisotropy Probe polarization data, finding a rotation angle $\beta_0=(0.34\pm0.09)^\circ$ at a $3.6\,\sigma$ preference~\cite{Eskilt:2022:biref-const}. 
Combining the Planck and Atacama Cosmology Telescope data yields the current tightest constraint on the birefringence angle as $(0.277\pm0.057)^\circ$, which is non-zero with a statistical significance of $4.8\,\sigma$ \cite{Eskilt:2026}. 
A number of studies have also investigated the effects of cosmic birefringence on the shape of the CMB angular power spectra \cite{Liu:2006:biref-time-evolve,Gubitosi:2014:biref-time,Lee:2013:biref,Sherwin:2021:biref,Nakatsuka:2022,Naokawa:2023,Carralot:2026,Yin&Komatsu:2026,Kamionkowski&Caldwell:2026}. However, the signal is consistent with a constant rotation \cite{Namikawa:2025:Planck,Salvatore:2025:biref} and no features specific to cosmic birefringence arising from the dynamics of ALPs have been found yet~\cite{Gruppuso:2016nhj,P16:rot,Eskilt:2023:EDE,Namikawa:2025:Planck,Yin:2025f:biref,Yin:2026:biref,Salvatore:2025:biref}. Further independent tests of the cosmic-birefringence signal constitute a natural and important step toward establishing evidence for new physics beyond both the standard spatially flat $\Lambda$ Cold Dark Matter ($\Lambda$CDM) cosmological model and the Standard Model of particle physics.

% CMB-BAO tension
Another potential tension within the standard $\Lambda$CDM model has emerged from recent baryon acoustic oscillation (BAO) measurements~\cite{DESI:2025:DR2,SPT-3G:2025}. When combined with CMB data, these measurements favor a value of the summed neutrino mass that lies below the minimum allowed by neutrino oscillation experiments~\cite{DESI:2025:neutrino,Craig:2024:neutrino,Green:2024:neutrino,Elbers:2024:neutrino,Lynch:2025:neutrino-tension,Allali:2025yvp,Cozzumbo:2025ewt}. Within the standard cosmological framework, a proposed resolution of this tension involves reinterpreting the discrepancy as a shift in the CMB optical depth, $\tau$, as constrained by Planck~\cite{Giare:2023:tau,Loverde:2024nfi,Allali:2025wwi,Jhaveri:2025:tau,Sailer:2025:tau,Allali:2025yvp}. 
However, a large value of $\tau\sim 0.09$ is not favored by the Planck low-$\l$ polarization data and increasing $\tau$ within $\Lambda$CDM has not succeeded in resolving the discrepancy. 

% cosmic birefringence and CMB-BAO tension
In Ref.~\cite{Namikawa:2025doa} it was pointed out that cosmic birefringence can resolve the ``$\tau$ tension'' if the birefringence angle varies significantly during the epoch of reionization. Because the observed polarization signal is integrated along the line of sight, substantial variations in the rotation angle during reionization cause phase mixing of the polarization signal. Consequently, the polarization amplitude is suppressed through averaging over different rotation phases.
A recent study in Ref.~\cite{Naokawa:2024xhn} showed that cosmic birefringence induced by axion-like particles (ALPs) can produce a large rotation angle owing to the ``$n\pi$-phase ambiguity'' in recent measurements. Specifically, the rotation angle that fits the data is degenerate with $\beta_0+180^\circ \times n$ for $n\in\mathbb{Z}$. Hereafter, we refer to this model as the $n\pi$ model. If $n\neq 0$, the model can accommodate a larger value of $\tau$, thereby alleviating the tension between the CMB and DESI results while remaining consistent with the birefringence angle constrained by the $EB$ power spectrum \cite{Namikawa:2025doa}.
Another possible mechanism for generating a large rotation angle arises when energy from a dark matter component is transferred to the ALPs during or before the epoch of reionization, but after recombination. Hereafter, we refer to this scenario as the energy-injection model. As pointed out in Ref.~\cite{Graham:2025dqn}, such energy injection enhances the ALP field amplitude well after recombination and produces a large rotation angle during reionization. Ref.~\cite{Sirletti:2026pgv} further explored the parameter space of the energy-injection model in which the low-$\ell$ $E$-mode power spectrum is suppressed with $\tau\simeq 0.09$ and the Planck low-$\l$ polarization data are well-fitted.

% This paper
In this work, we test the above scenarios using the radio galaxy catalog presented in Ref.~\cite{Battye-Jackson-Browne:2025}, hereafter referred to as BJB26. Radio galaxies have been used to constrain cosmic birefringence under the empirical relation \cite{Gardner_1963, Gardner_1966} that the jet direction is, on average, perpendicular to the polarization angle of a radio galaxy \cite{Carroll:1989:rot,Carroll:1997:radio,Eisenstein:1997sa,Leahy:1997wj,Carroll:1998:DE,Battye-Jackson-Browne:2025}.
By using this ``standard cross'' \cite{Naokawa:2025:DEprofile} of radio galaxies with known redshifts, we can constrain cosmic birefringence at much lower redshifts than those corresponding to the epochs of reionization and recombination. At such low redshifts, CMB polarization is not efficiently generated and therefore cannot tightly constrain cosmic birefringence. Radio galaxy observations constrain cosmic-birefringence models independently of CMB data and provide a unique opportunity to test the dark-energy profile proposed in Ref.~\cite{Naokawa:2025:DEprofile}.
The BJB26 radio galaxy data yield a constraint of $\beta=(0.2\pm 1.0)^\circ$ on cosmic birefringence when the X-band data are used to determine the jet position angle~\cite{Battye-Jackson-Browne:2025}. Because the rotation angle predicted by the models described above can vary significantly at low redshift, reaching values from sub-degree scales to a few tens of degrees at $z\lesssim 10$ depending on the model parameters, the BJB26 data are sensitive to these large-birefringence scenarios.

% Structure
This paper is organized as follows. In Sec.~\ref{sec:theory}, we review cosmic birefringence induced by ALPs. In Sec.~\ref{sec:data}, we describe the radio galaxy catalog and the method used to estimate the cosmic birefringence angle. In Sec.~\ref{sec:results}, we present constraints on cosmic-birefringence scenarios using data from radio galaxy observations. Finally, Sec.~\ref{sec:discussion} is devoted to a summary of our results and conclusions.

%////////////////////////////////////////%
\section{Cosmic birefringence} \label{sec:theory}
%////////////////////////////////////////%

Here, we briefly review cosmic birefringence induced by ALPs. Cosmic birefringence can arise from a pseudoscalar field $\phi$, such as ALPs, coupled to the electromagnetic field via a Chern–Simons interaction:
%--------------------------------------------------%
\al{
    \mathcal{L}\supset -\frac{1}{4}g\phi F_{\mu\nu}\tilde{F}^{\mu\nu}
    \,,
} 
%--------------------------------------------------%
where $g$ is the coupling constant, $F_{\mu\nu}$ is the electromagnetic field tensor, and $\tilde{F}^{\mu\nu}$ is its dual. This mechanism has been widely explored in various cosmological contexts, including ALPs as dark energy candidates~\cite{Carroll:1998:DE,Liu:2006:biref-time-evolve,Panda:2010,Fujita:2020:forecast,Fujita:2020:biref,Choi:2021aze,Obata:2021,Gasparotto:2022uqo,Galaverni:2023}, early dark energy scenarios~\cite{Fujita:2020:biref,Murai:2022:EDE,Eskilt:2023:EDE,Kochappan:2024:biref,Yin:2025f:biref,Yin:2025f:biref}, and axion dark matter models~\cite{Finelli:2009,Sigl:2018:biref-sup,Liu:2016:AxionDM,Fedderke:2019:biref,Zhang:2024dmi,Shao:2025jhp}. Additional sources proposed include topological defects~\cite{Takahashi:2020tqv,Kitajima:2022jzz,Jain:2022jrp,Gonzalez:2022mcx,Lee:2025:biref-DW} and possible quantum gravity effects~\cite{Myers:2003fd,Balaji:2003sw,Arvanitaki:2009fg}.

% time-evolving beta
When cosmic birefringence is sourced by an ALP field, the rotation angle becomes time-dependent, and its evolution depends on the mass of ALPs, $m_\phi$. The net rotation angle for a photon observed today and emitted at conformal time $\eta$ is given by \cite{Carroll:1989:rot,Carroll&Field:1991,Harari:1992}
%--------------------------------------------------%
\al{
    \beta(\eta) 
    &= \frac{g}{2}[\phi(\eta_0)-\phi(\eta)] 
    = \frac{g\phi_{\rm ini}}{2}\frac{\phi(\eta_0)-\phi(\eta)}{\phi_{\rm ini}}
    \,, \label{Eq:beta}
}
%--------------------------------------------------%
where $\eta_0$ is the present conformal time and $\phi_{\rm ini}$ is the initial value of $\phi$. 
In this paper, we choose $g\phi_{\rm ini}/2$ so that the cosmic birefringence angle at the recombination provides $\beta_0=0.34^\circ$. 
For a single ALP without interactions with other components in the Universe, the evolution of $\phi$ is governed by
%--------------------------------------------------%
\al{
    \phi'' + 2{\cal H}\phi' + a^2V_{,\phi}(\phi) = 0
    \,, \label{Eq:phi-EoM}
}
%--------------------------------------------------%
where $a$ is the scale factor, prime is the derivative with respect to the conformal time, ${\cal H}=a'/a$, and $V_{,\phi}(\phi)$ is the partial derivative of the potential $V(\phi)$ with respect to $\phi$. In the following, we assume a cosine potential $V(\phi)=m_\phi^2f_\phi^2[1-\cos(\phi/f_\phi)]$ where $f_\phi$ is the decay constant. 

Some of the mass ranges are already excluded for the $n\pi$ model if $n\not=0$. 
If $m_\phi\gg 10^{-28}\,$eV, even if $n=0$, ALPs evolve during the recombination and the CMB power spectra are modified significantly \cite{Finelli:2009,Fedderke:2019:biref,Zhang:2024dmi,Murai:2024yul,Yin&Komatsu:2026}. 
However, the high-$\l$ Planck polarization data favors a constant rotation that excludes $m_{\phi}\agt 10^{-28}\,$eV \cite{Namikawa:2025:Planck}. 
Heavy mass ALPs $m_\phi\gg 10^{-28}\,$eV with $n\not=0$ are also not favored by the latest measurements of the anisotropic cosmic birefringence \cite{Namikawa:2020:biref,SPT:2020:biref,BK-LoS:2023,Zagatti:2024}. 
If $n\not=0$, even with $m_{\phi}\gg 10^{-30}\,$eV, the rotation angle changes significantly during the recombination epoch and the shape of the CMB power spectra is significantly modified compared to the constant rotation case \cite{Naokawa:2024xhn}. Thus, we limit our mass range to $m_{\phi}\leq 10^{-30}\,$eV. 
In contrast, if $m_\phi\ll 10^{-31}\,$eV with $n\not=0$, the late-time birefringence angle remains significant even at low redshifts and the amplitude of cosmic birefringence angle becomes more than $\mC{O}(1^\circ)$ that is already disfavored by the radio galaxy birefringence measurement \cite{Battye-Jackson-Browne:2025}. Thus, for the $n\pi$ model, we consider the ALP masses in the range $10^{-31}\,{\rm eV} \lesssim m_{\phi} \lesssim 10^{-30}\,{\rm eV}$. 

In the energy-injection model, we consider an interaction between ALPs and the cold dark matter (CDM). If ALP interacts with other components, the right-hand-side of Eq.~\eqref{Eq:phi-EoM} becomes nonzero. Assuming that a tiny fraction of the energy of the CDM transfers into ALPs \cite{Graham:2025dqn,Sirletti:2026pgv}, we consider the simple model where the continuity equations for the ALPs and CDM are described as (e.g. \cite{Wang:2016lxa}):
\al{
    \rho'_{\phi} + 3{\cal H}(\rho_{\phi}+p_{\phi}) 
    &= {\cal H}\xi\rho_{\rm c}
    \,, \\
    \rho'_{\rm c} + 3{\cal H}\rho_{\rm c} 
    &= -{\cal H}\xi\rho_{\rm c}
    \,, \label{Eq:rho_c:evolve}
}
where $\rho_{\phi}$ and $p_\phi$ are the energy density and pressure of the ALPs, respectively, $\rho_{\rm c}$ is the energy density of CDM, and $\xi$ is the efficiency of the energy injection. We rewrite the above equation in terms of $\phi$ and obtain \footnote{The right-hand side of Eq.~\eqref{Eq:phi-EoM:inject} diverges if $\phi'=0$. However, $\phi'$ does not vanish during the energy injection.}
%--------------------------------------------------%
\al{
    \phi'' + 2{\cal H}\phi' + a^2V_{,\phi}(\phi) = \frac{a^2{\cal H}\xi\rho_{\rm c}}{\phi'}
    \,. \label{Eq:phi-EoM:inject}
}
%--------------------------------------------------%
In our analysis, we assume that $\xi(z)$ is constant, $\xi$, when $z_{\rm end}\leq z\leq z_{\rm start}$ and is zero otherwise. If we inject too much energy into ALPs by increasing $\xi$, $\phi$ crosses over the peak of the cosine potential and cannot oscillate around $\phi=0$ after the energy injection. Thus, there is a critical value of the energy transfer efficiency, $\xi=\xi_{\rm crit}$ \cite{Sirletti:2026pgv}. 

We assume the initial condition that satisfies $1/\tilde{f}_{\phi}\equiv \phi_{\rm ini}/f_\phi=0.01\ll 1$ and $\phi'_{\rm ini}=0$ so that the ALP field value becomes large enough to realize a large rotation angle after the energy injection period. In this case, the initial ALP energy density is $\rho_{\phi,\rm ini}\simeq m_\phi^2\phi_{\rm ini}^2/2$.
We assume that the CDM energy density remains close to that in the standard $\Lambda$CDM model and evolves as $\rho_{\rm c,0}/a^3$ where $\rho_{\rm c,0}$ is the present energy density of CDM. We also assume that ALP behaves as a spectator field and the background evolution remains the same as that in the standard $\Lambda$CDM model. 
Denoting $x\equiv\phi/\phi_{\rm ini}$, the evolution equation is rewritten as
%--------------------------------------------------%
\al{
    x'' + 2{\cal H}x' + a^2V_{,x}(x) = \frac{{\cal H}}{ax'}\frac{m_\phi^2}{2}\xi\frac{\rho_{\rm c,0}}{\rho_{\phi,\rm ini}}
    \,, \label{Eq:x-EoM:inject}
}
%--------------------------------------------------%
where $V(x)=m_\phi^2\tilde{f}_\phi^2[1-\cos(x/\tilde{f}_\phi)]$. \footnote{The right-hand side of Eq.~\eqref{Eq:x-EoM:inject} changes slightly if we use the exact solution for $\rho_{\rm c}$ from Eq.~\eqref{Eq:rho_c:evolve}, which depends on $\xi$. However, this correction is of order $\mC{O}(\xi^2)$, and its effect on the right-hand side can be made arbitrarily small by choosing a sufficiently small value of $\xi$ while $\tilde{\xi}$ is sufficient to produce a large rotation angle \cite{Sirletti:2026pgv}.}
For a given mass and $\tilde{f}_\phi$, the evolution of $x=\phi/\phi_{\rm ini}$ depends only on the rescaled energy-transfer efficiency
\al{
    \tilde{\xi} \equiv \xi \frac{\rho_{\rm c,0}}{\rho_{\phi,\rm ini}}
    = \frac{\xi}{F}
    \,,
}
where $F\equiv\rho_{\phi,\rm ini}/\rho_{\rm c,0}$. Since the measured birefringence angle is determined by $\tilde{\xi}$, we constrain $\tilde{\xi}$ rather than $\xi$ in our analysis.

Let us briefly comment on the possible values of $\tilde{\xi}=\xi/F$. As discussed in Ref.~\cite{Sirletti:2026pgv}, the parameters $\xi$ and $F$ cannot be chosen freely because they are subject to the following three conditions. First, the CDM energy density should remain close to that in the standard $\Lambda$CDM model. Second, the ALP should behave as a spectator field. Third, $\xi$ must satisfy $\xi<\xi_{\rm crit}$ which is equivalent to $\tilde{\xi}<\tilde{\xi}_{\rm crit}=\xi_{\rm crit}/F$. For the mass range of interest, however, we can choose $\xi$ and $F$ so as to satisfy all three conditions while allowing $\tilde{\xi}$ to approach $\tilde{\xi}_{\rm crit}$ \cite{Sirletti:2026pgv}. Thus, in the following analysis, we vary $\tilde{\xi}$ between $0$ and $\tilde{\xi}_{\rm crit}$. 

\begin{figure}[t]
\bc
\includegraphics[width=8.5cm,clip]{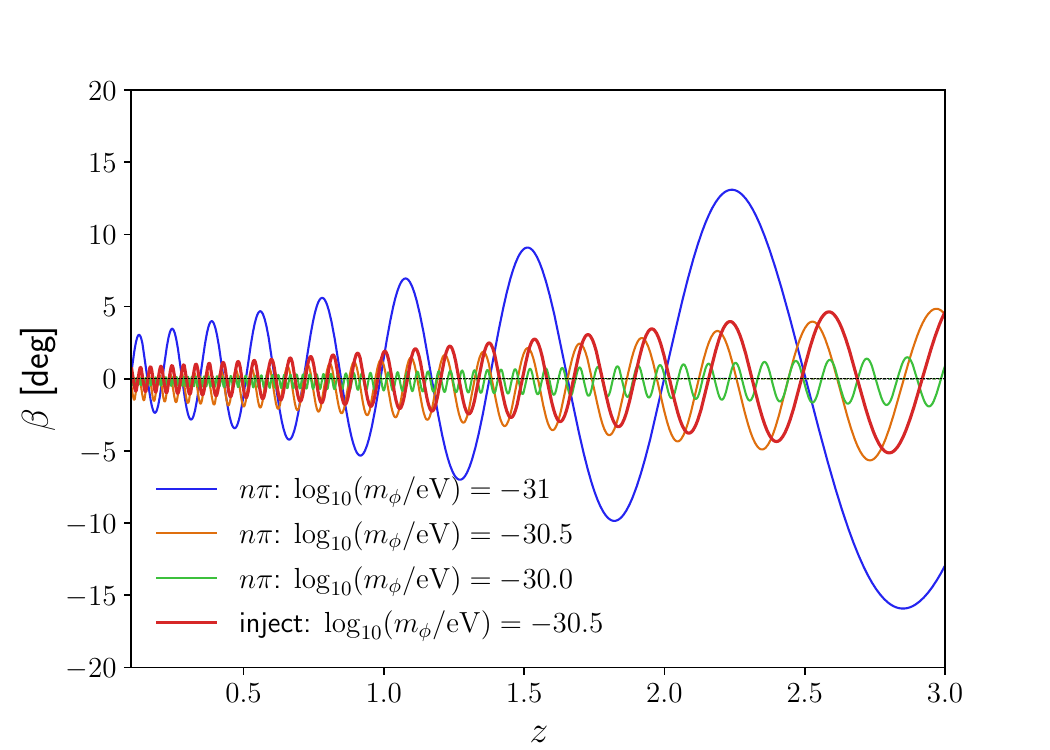}
\caption{
Cosmic birefringence angle as a function of redshift for specific cases of the logarithmic ALP masses, $\log_{10}(m_\phi/{\rm eV})$, for the $n\pi$ and energy-injection models. For the $n\pi$ model, we assume $n=1$ for all cases. If $n=0$ with the same masses, the birefringence angle at $z\leq 3$ is much smaller than $0.3^\circ$ and is almost overlapped with the $\beta=0^\circ$ horizontal line in the figure. For the energy-injection model, we assume the rescaled energy injection efficiency, $\tilde{\xi}=10$, and the start and end of the energy injection period in redshift, $z_{\rm start}=60$ and $z_{\rm end}=50$. 
}
\label{fig:phi}
\ec
\end{figure}

Figure~\ref{fig:phi} shows the cosmic birefringence angle as a function of redshift for the $n\pi$ and energy-injection models for several representative parameter sets. For the $n\pi$ model, we assume $n=1$ for all of the cases with varying $m_\phi$. If $n=0$, the rotation angle becomes small and overlaps with the $\beta=0^\circ$ horizontal line in the plot. If $n$ increases, the lower mass cases significantly deviate from the no-rotation case. In the energy injection model, if we set $\tilde{\xi}\simeq 10$, the rotation angle becomes large. This situation depends strongly on the initial condition of the $\phi$ field. If $\phi$ starts sufficiently close to the minimum of the potential, then even with a relatively large energy injection, the ALP field does not cross the potential barrier. In that case, the value of $\tilde{\xi}$ required to suppress the low-$\l$ CMB $EE$ does not necessarily have to be as large as $\tilde{\xi}_{\rm crit}$. In this sense, using a quadratic potential instead of the cosine potential should also work for a large rotation angle after the recombination by the energy injection. 
To constrain these models, we need to observe the rotation angle with its uncertainties at the order of degrees at each redshift.

%////////////////////////////////////////%
\section{Birefringence angle in redshift bins} \label{sec:data}
%////////////////////////////////////////%

\begin{figure}[t!]
  \centering
  \includegraphics[width=0.8\columnwidth]{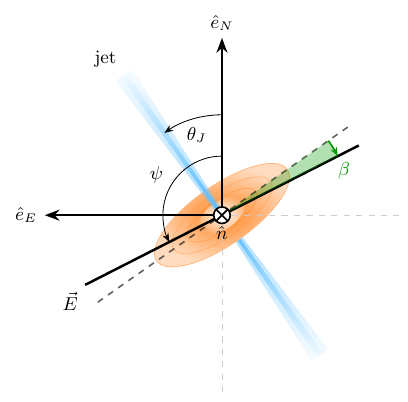}
  \caption{
    Geometry of the birefringence measurement on the sky plane, as seen by the observer. The unit vectors $\hat{e}_N$ and $\hat{e}_E$ point toward the North and East, and $\hat{n}$ is the line-of-sight direction, pointing from the observer to the source (into the page). Since the sky is observed from inside the celestial sphere, East lies to the left of North and the triad $(\hat{e}_N, \hat{e}_E, \hat{n})$ is left-handed, i.e., $\hat{e}_N \times \hat{e}_E = -\hat{n}$, where $-\hat{n}$ is the photon propagation direction. Position angles are measured from North through East (IAU convention): $\theta_J$ is the position angle of the jet axis and $\psi$ that of the observed polarization direction $\vec{E}$ (solid line). The dashed line indicates the intrinsic polarization direction, expected to be perpendicular to the jet, $\theta_J - 90^\circ$. Cosmic birefringence rotates the polarization plane by the angle $\beta_J = \theta_J - \psi - 90^\circ$ (green shaded region), with $\beta_J > 0$ corresponding to a clockwise rotation on the sky, as in the CMB convention. 
  }
  \label{fig:sky_angles}
\end{figure}

\begin{figure*}[p]
\centering
\includegraphics[width=\textwidth,clip]
  {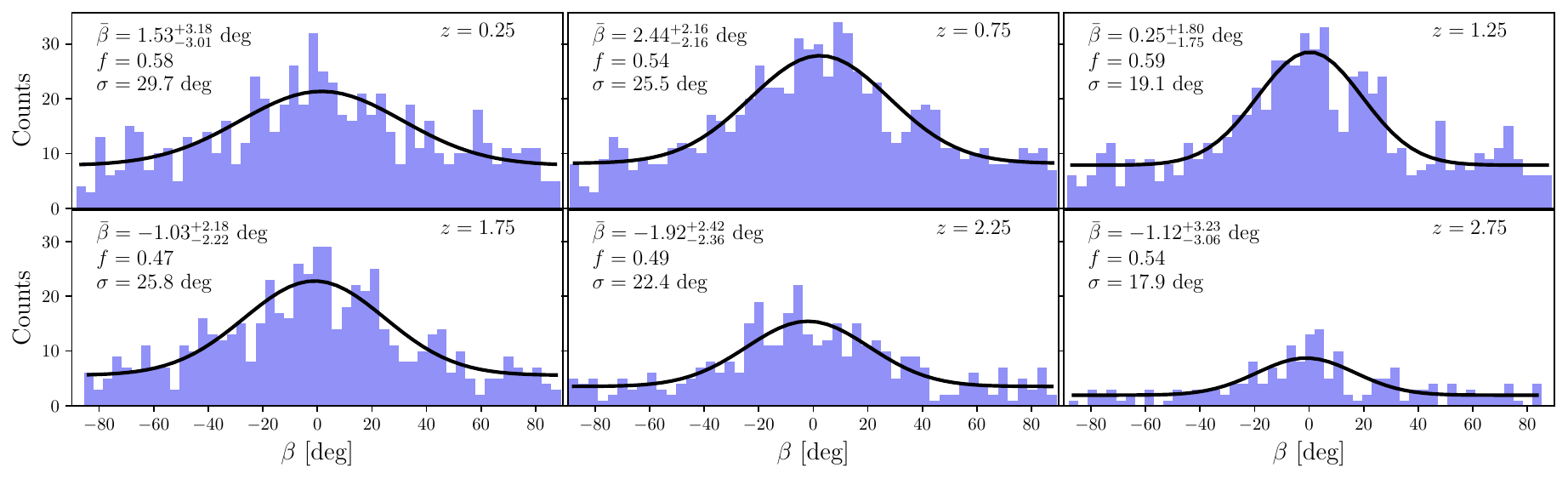}
\caption{
Histogram of the cosmic birefringence angles
at each redshift bin with $\Delta z=0.5$.
}
\label{fig:hist:0.5}

\vspace{1cm}

\includegraphics[width=\textwidth,clip]
  {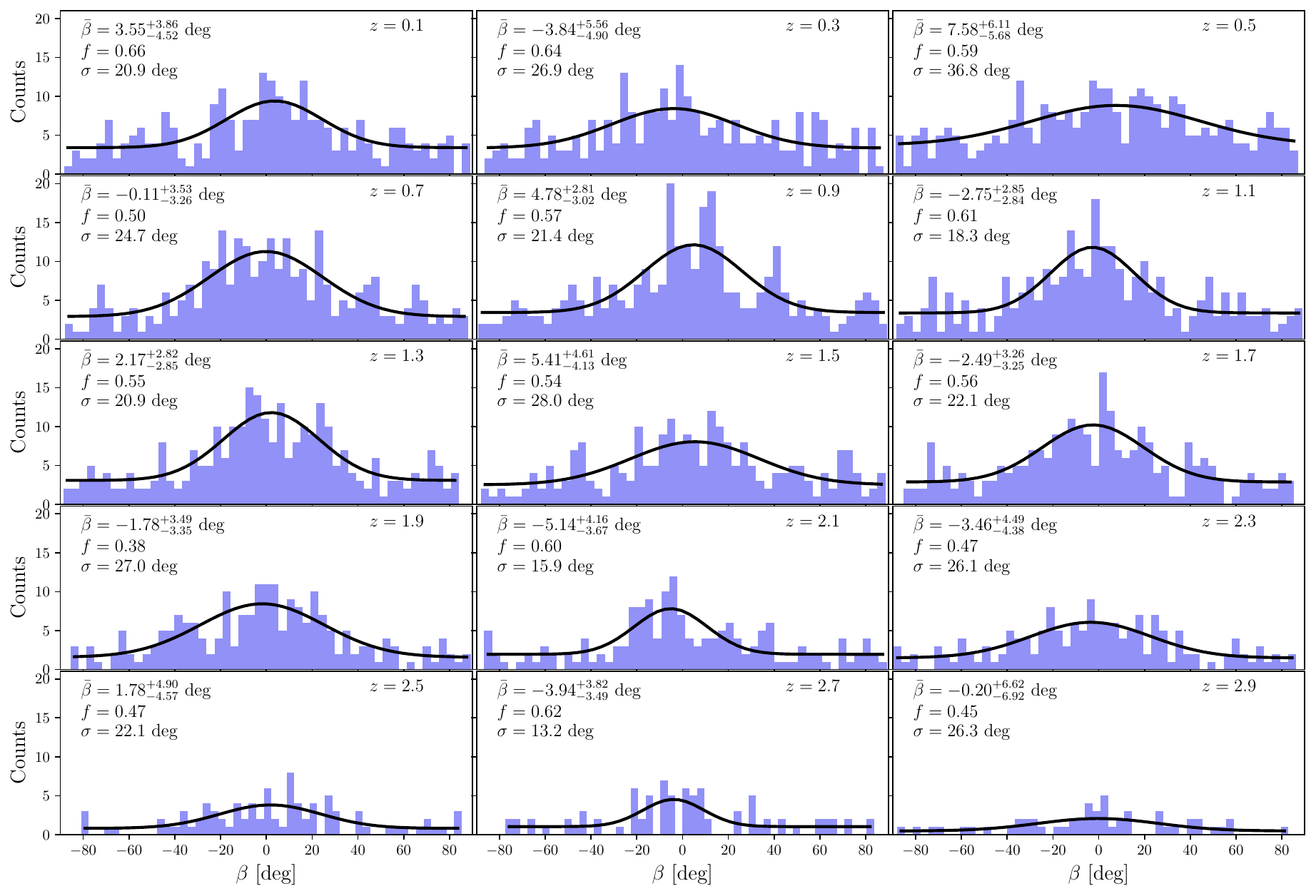}
\caption{
Same as Fig.~\ref{fig:hist:0.5} but for
$\Delta z=0.2$.
}
\label{fig:hist:0.2}
\end{figure*}

\begin{figure*}[t]
\centering
\includegraphics[width=16cm]{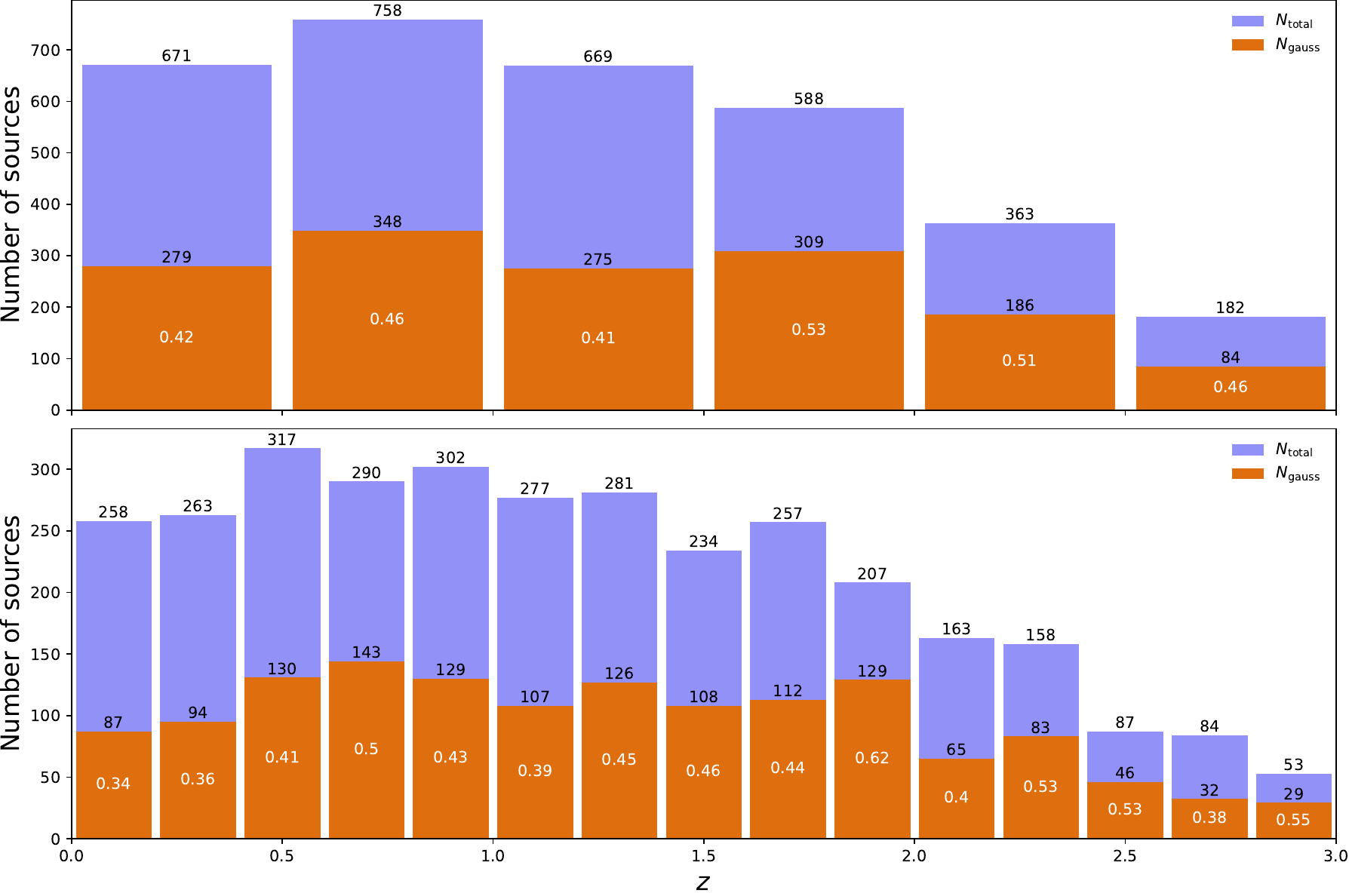}
\caption{
Total number of radio galaxy sources per redshift bin, $N_{\rm total}$, and estimated number of sources that belong to Gaussian components, $N_{\rm gauss}$. The numbers shown at each bin are $N_{\rm total}$, $N_{\rm gauss}$, and the fraction, $N_{\rm gauss}/N_{\rm total}$, from top to bottom, respectively.}
\label{fig:num}
\end{figure*}

\vspace{1cm}

\begin{figure*}[t]
\includegraphics[width=17cm]{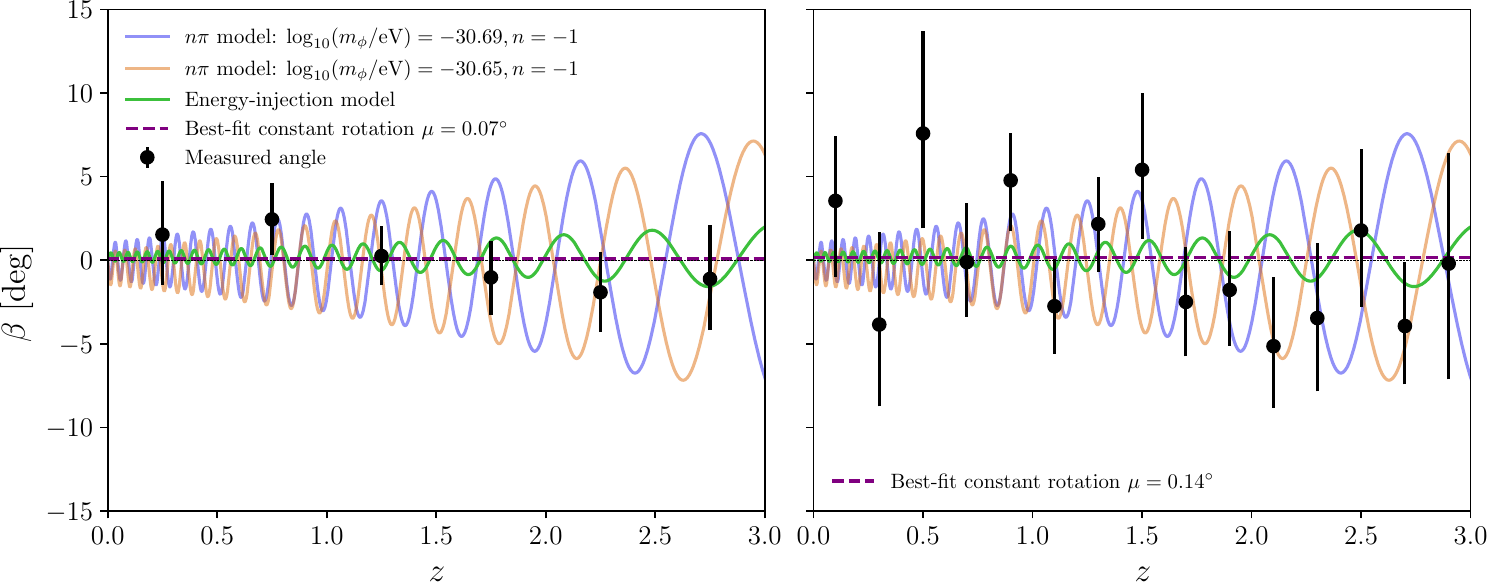}
\caption{
Measured cosmic birefringence angle $\bar{\beta}$ at each redshift bin obtained from Figs.~\ref{fig:hist:0.5} and \ref{fig:hist:0.2}, shown as black points with error bars. For comparison, we also show representative theoretical predictions for the $n\pi$ model with $n=-1$ and for the energy-injection model with $\tilde{\xi}=13$, $z_{\rm start}=60$, $z_{\rm end}=50$, and $\log_{10}(m_\phi/{\rm eV})=-30.6$, together with the best-fit constant rotation obtained by fitting the birefringence angles in each sample using Eq.~\eqref{Eq:fit}. For the $n\pi$ model, we show the best-fit model, $\log_{10}(m_\phi/{\rm eV}) = -30.69$ with $n = -1$, obtained from the analysis with redshift binning, and a model excluded at more than $2\sigma$, $\log_{10}(m_\phi/{\rm eV}) = -30.65$ with $n = -1$.
}
\label{fig:beta-z}
\end{figure*}

This section describes how we obtain the birefringence angle for each radio galaxy sample from the catalog developed by BJB26. Specifically, we calculate the constraints on the birefringence angle as a function of redshift. 

For each radio galaxy in the catalog, the structural position angles of the jet measured in the S, C, X, and U bands, $\theta_S$, $\theta_C$, $\theta_X$, and $\theta_U$, are provided, together with the polarization position angle $\psi$ and redshift $z$. We derive the birefringence angle in each band as $\beta_J=\theta_J-\psi-90^\circ$, where $J=S,C,X,U$, taking into account that $\beta\in(-90^\circ,90^\circ)$. Note that the position angles of the jet and polarization are measured following the IAU convention, from North through East, which is opposite to the convention commonly adopted in CMB observations \cite{Battye-Jackson-Browne:2025}. Figure~\ref{fig:sky_angles} shows the geometry underlying this estimator.

For each sample, if we have $\beta_J$ for that sample at multiple bands, we simply average them to create a single cosmic birefringence angle for that sample. We then obtain a set of measured birefringence angle data, $\beta_i$, where $i$ is the index of the samples. 
Note that we neglect Faraday rotation in this analysis since its correction is negligible, compared to the statistical uncertainty of the birefringence angle measurement as shown in BJB26. 

To see the redshift dependence of the birefringence angle, we divide the samples into bins based on their redshift. We use samples whose redshift is in the range between $0$ and $3$. 
We consider two cases of the bin width; $\Delta z=0.2$ and $0.5$. 
At each redshift bin, we use MCMC to fit the histogram of $\beta_i$ from the subsamples with a sum of the uniform and Gaussian distributions proposed by BJB26 where we have three parameters to be fit with data; the fraction of the uniform distribution $f$, and the mean and standard deviation of the Gaussian distribution, $\bar{\beta}$ and $\sigma$, respectively. We then obtain the best-fit value of the mean of the Gaussian distribution, $\bar{\beta}$, and its posterior distribution at each redshift bin. We use the posterior distribution to evaluate the error of $\bar{\beta}$. 

Figures~\ref{fig:hist:0.5} and \ref{fig:hist:0.2} show the fitting results and histogram at each redshift bin. At each redshift bin, we divide the samples of the cosmic birefringence angle into $50$ bins to create a histogram.
We check that each histogram is in good agreement with the fitted curve. 
Figure~\ref{fig:num} also shows the total number of radio galaxies, the estimated number of sources that belong to the Gaussian components, and the fraction of the Gaussian components to the total number of sources. 

Figure~\ref{fig:beta-z} shows $\beta(z)$ at each redshift bin with $\Delta z=0.5$ and $0.2$ cases. We also show an example of the theoretical prediction with specific values of $n$ and $\log_{10}(m_{\phi}/{\rm eV})$. For both redshift binning schemes, the rotation angles in the low-redshift bins ($z<1.5$) are slightly positive, whereas those in the high-redshift bins ($z>1.5$) are slightly negative. Setting $\Delta z=0.2$ reduces the number of sources per bin, which increases the statistical uncertainties and also increases the number of free parameters to be fitted. We therefore adopt only $\Delta z=0.5$ for the redshift binning in the following analysis. 

%////////////////////////////////////////%
\section{Constraints on cosmic birefringence models} \label{sec:results}
%////////////////////////////////////////%

\begin{figure*}[t]
\centering
\includegraphics[width=0.49\textwidth]{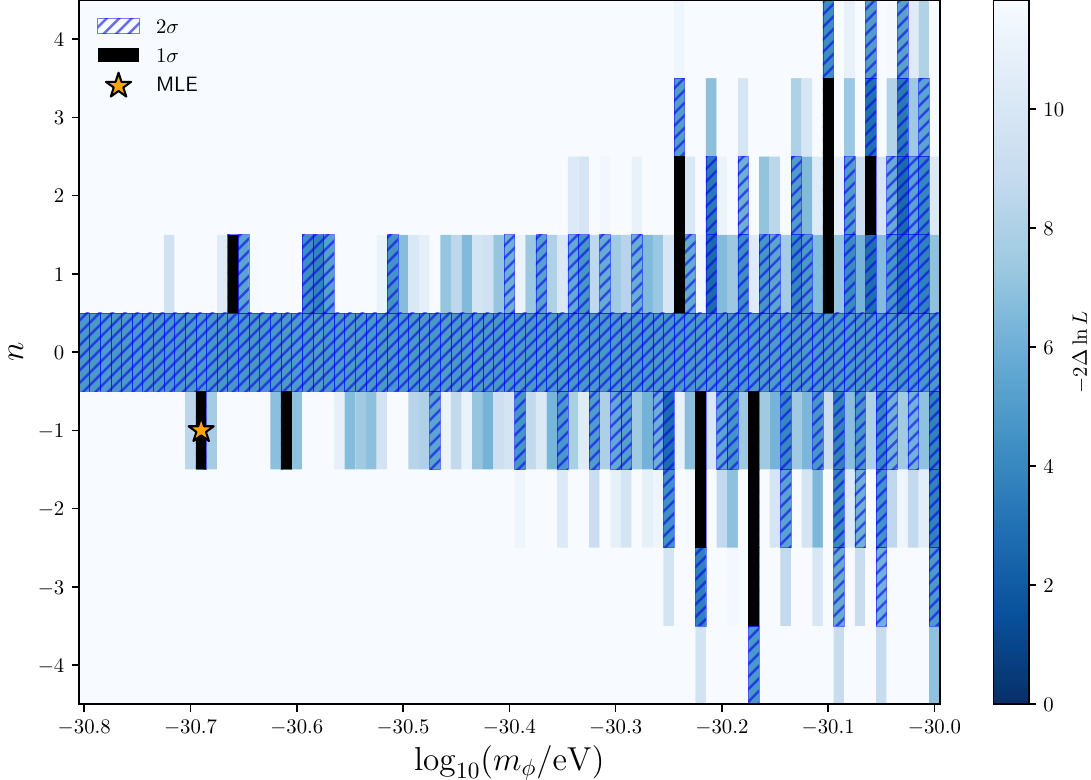}\hfill
\includegraphics[width=0.49\textwidth]{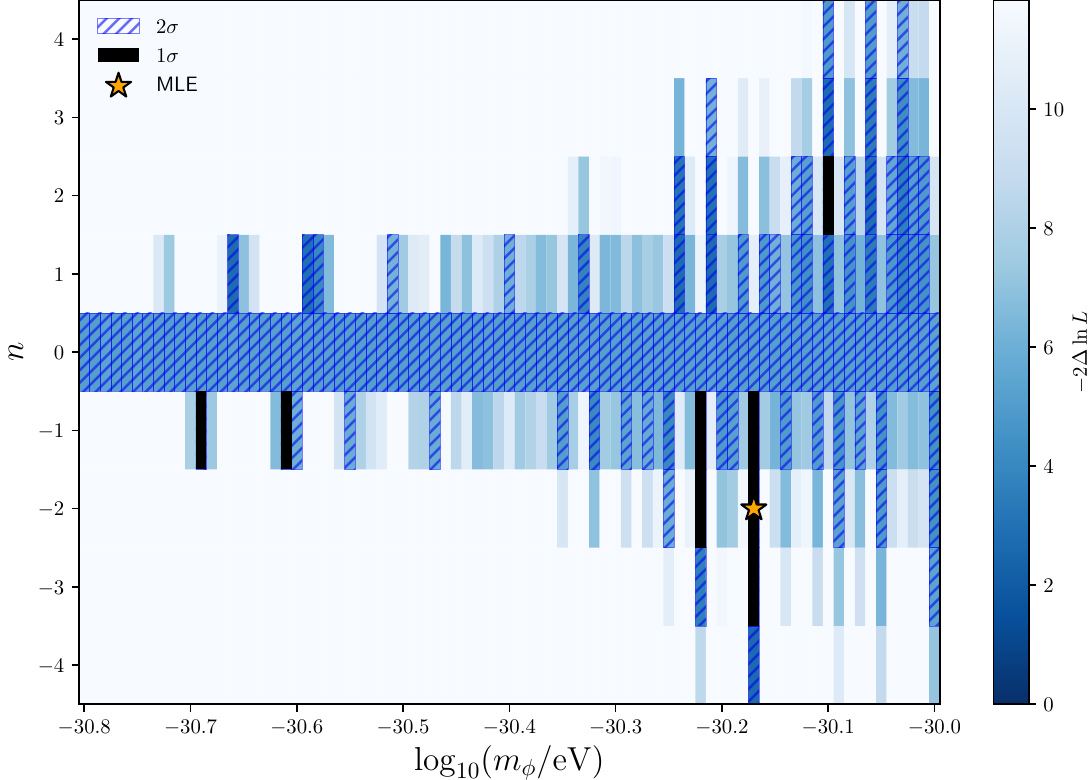}
\caption{
Constraints on the ALP mass and ambiguity parameter $n$ of the $n\pi$ cosmic birefringence model with redshift binning (left) and without redshift binning (right). For each ALP mass and $n$, we compute the likelihood value maximized by varying $\sigma_b$ and $f_b$. The plot shows the profile log-likelihood difference, $\Delta\ln{\cal L}(m_\phi,n)=\max_{p}\ln{\cal L}(m_\phi,n,p)-\max_{m_\phi,n,p}\ln{\cal L}(m_\phi,n,p)$, together with the corresponding $1\sigma$ and $2\sigma$ intervals relative to the best-fit value denoted as the orange star. 
}
\label{fig:const-npi-2d}
\end{figure*}

\begin{figure*}[t]
\includegraphics[width=0.49\textwidth]{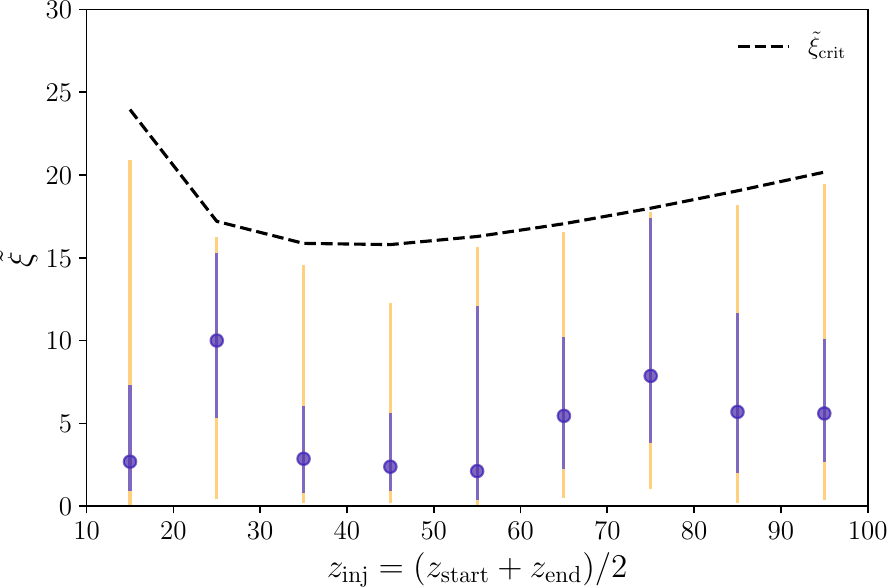}\hfill
\includegraphics[width=0.49\textwidth]{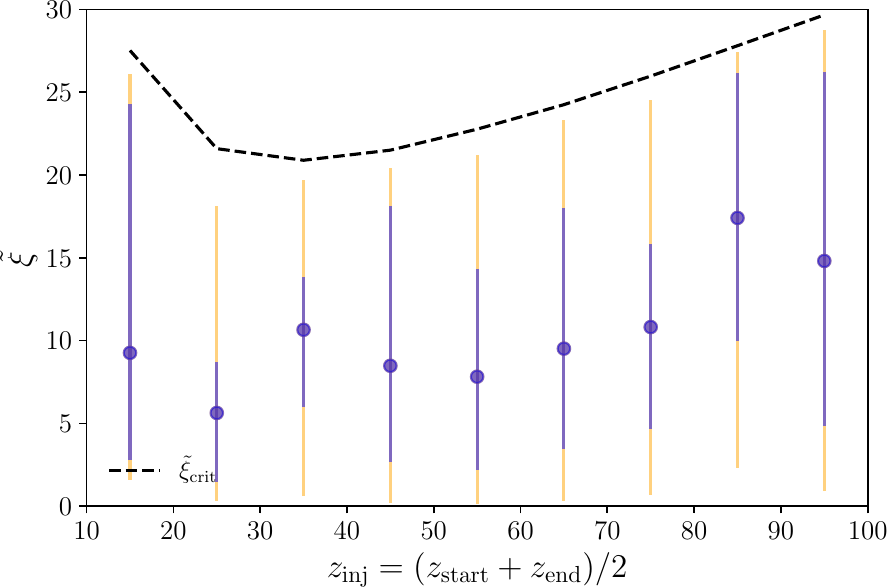}
\caption{
Constraints on the rescaled energy injection efficiency, $\tilde{\xi}$, for the energy injection model with $1$ and $2\,\sigma$ error bars in blue and orange, respectively. The black dashed curve shows the critical value of the energy injection efficiency, $\tilde{\xi}_{\rm crit}$, beyond which the ALP dynamics becomes very different and the energy injection model cannot alleviate the CMB-BAO tension \cite{Sirletti:2026pgv}. We fit the data with $\tilde{\xi}$, $f_b$, and $\sigma_b$, for given ALP mass and energy injection period, $z_{\rm start}$ and $z_{\rm end}$. The left and right panels show the cases with $\log_{10}(m_{\phi}/{\rm eV})=-30.5$ and $\log_{10}(m_{\phi}/{\rm eV})=-30.6$, respectively. 
}
\label{fig:const-xi}
\end{figure*}

We next constrain the cosmic-birefringence models that suppress the large-scale $E$-mode power spectrum and thereby alleviate the CMB–BAO tension. 
Our procedure is as follows. 

To constrain the $n\pi$ model, for each ALP mass $m_\phi$ and the phase parameter $n$, we compute a theoretical template for the cosmic birefringence angle as a function of redshift, $\beta^{\rm th}(z)$. We then evaluate the likelihood
\al{
    \ln\mC{L} &= \sum_i\ln\bigg[(1-f(z_i))N(\beta_i;\beta^{\rm th}(z_i),\sigma(z_i)) 
    \notag \\
    &\qquad+ f(z_i)U(\beta_i;-90^\circ,90^\circ)\bigg]
    \,, \label{Eq:fit}
}
where $N(\beta;\mu,\sigma)$ denotes a normal distribution with mean $\mu$ and standard deviation $\sigma$, and $U(\beta;-90^\circ,90^\circ)$ denotes a uniform distribution over $-90^\circ\leq\beta\leq90^\circ$. To account for the redshift dependence of $f$ and $\sigma$, we assume that both quantities can vary with redshift bins but are constant within each redshift bin. We obtain the maximum likelihood value and the best-fit parameters, $f_b$, and $\sigma_b$, with $b=1,\ldots,N_z$ that maximizes the likelihood for a given $m_\phi$ and $n$. 

To constrain the energy-injection model, we compute the theoretical template for a given ALP mass and energy-injection period. We then compute the posterior distribution of the parameters, $\tilde{\xi}$, $f_b$, and $\sigma_b$. 

As a consistency check, we also fit the data assuming a redshift-independent rotation angle by replacing $\beta^{\rm th}(z_i)$ with a constant $\mu$. We do not apply the redshift binning. In this limit, the analysis closely resembles that of BJB26. We obtain $\mu=(0.4\pm0.9)^\circ$, which is consistent with the BJB26 result, $\mu=(0.2\pm1.0)^\circ$, obtained using the X-band data.\footnote{The small difference between the two results arises from our inclusion of birefringence-angle measurements from all four bands.}

Figure~\ref{fig:const-npi-2d} shows the constraints on the ALP mass and $n$, where we plot the log-likelihood difference,
\al{
    \Delta\ln\mC{L}(m_\phi,n) = \max_{p}\ln\mC{L}(m_\phi,n,p) - \max_{m_\phi,n,p}\ln\mC{L}(m_\phi,n,p)
    \,,
}
where $p$ collectively denotes the remaining model parameters. We show the one and two sigma confidence regions assuming that the above profile log-likelihood difference $-2\Delta\ln{\cal L}(m_\phi,n)$ follows Wilks' theorem with two degrees of freedom. We show both cases with and without the redshift binning. In the redshift binning case, we divide samples into six equally-spaced redshift bins. The redshift binning does not change the allowed region significantly. 
Nonzero values of $n$ are excluded at more than $2\,\sigma$ if $\log_{10}(m_\phi/{\rm eV})\alt -30.7$. For most ALP masses, large values of $|n|$ are excluded by the radio galaxy observation. 
The likelihood is slightly enhanced where $|n|=\mC{O}(1)$ at some masses. 
Using the Planck CMB $EE$ power spectrum data for the analysis done in Ref.~\cite{Namikawa:2025doa}, the $|n|=1$ case with $-31.0\alt\log_{10}(m_{\phi}/{\rm eV})\alt-30.0$ lies within the $2\sigma$ parameter region that allows for a larger value of $\tau$. Similarly, the $|n|=2$ case with $-30.6\alt\log_{10}(m_{\phi}/{\rm eV})\alt-30.0$ lies within the $2\sigma$ parameter region that allows for a larger value of $\tau$. The radio-galaxy observation thus narrows, but does not rule out, the viable parameter space of the $n\pi$ model obtained from the Planck low-$\ell$ EE data. 

Figure~\ref{fig:const-xi} shows the constraints on $\tilde{\xi}$ for the energy-injection model. We fix the ALP mass and the beginning and end of the energy-injection period, parametrized by the redshifts $z_{\rm start}$ and $z_{\rm end}$. We assume $z_{\rm start}-z_{\rm end}=10$. 
We adopt a uniform prior on $\tilde{\xi}$ over $[0, \tilde{\xi}_{\rm crit}]$ and show the $1\,\sigma$ and $2\,\sigma$ credible intervals of its marginalized posterior. From our analysis, the data are consistent with the no injection model but show a slight relative preference for energy injection compared to the no injection model. The preference for nonzero energy injection is only at most at the $\sim 2\sigma$ level and depends on the ALP mass and injection epoch.
The preference could be just a statistical fluke, and further measurements of radio galaxies are important to test the models. 
Note that the $2\,\sigma$ preferred range of $\tilde{\xi}$ obtained in this analysis overlaps with that obtained using the Planck low-$\ell$ EE data with a larger value of the CMB optical depth, $\tau=0.09$ \cite{Sirletti:2026pgv}. 

%////////////////////////////////////////%
\section{Summary and Discussion} \label{sec:discussion}
%////////////////////////////////////////%

% summary
We constrained the cosmic birefringence models with a large rotation angle during reionization, i.e., the $n\pi$-phase ambiguity and energy injection models that are motivated by resolving the CMB-BAO tension. 
We found that these models are not ruled out by the low-redshift observations of cosmic birefringence with radio galaxies. 
For the $n\pi$ model, the radio-galaxy data exclude the low-mass region $\log_{10}(m_\phi/{\rm eV}) \lesssim -30.7$ for $|n| \geq 1$, but part of the parameter region favored by the Planck low-$\ell$ $EE$ data remains allowed. 
For the energy injection model, the data are consistent with the no-injection model but show a slight preference for energy injection compared to the no-energy injection case. The $2\,\sigma$ preferred range of $\tilde{\xi}$ obtained in this analysis overlaps with that obtained using the Planck low-$\ell$ EE data with a larger value of the CMB optical depth, $\tau=0.09$ \cite{Sirletti:2026pgv}.
However, our samples are limited to $\sim 3000$ and there could still be several astrophysical uncertainties and statistical fluctuations. Further exploration of cosmic birefringence using a large number of astrophysical standard crosses at low redshift, such as radio galaxies, is therefore important for testing the cosmic birefringence solution to the CMB–BAO tension. 
Large surveys of polarized sources with Australian Square Kilometre Array Pathfinder (ASKAP)~\cite{Vanderwoude_2024} and Square Kilometre Array (SKA)~\cite{OSullivan:2026auw} will enable such tests.

% future
Future measurements of large-scale CMB polarization spectra by Cosmology Large Angular Scale Surveyor (CLASS) \cite{CLASS:2025} and LiteBIRD \cite{LiteBIRD,LiteBIRD:2025:biref} will offer a stringent test of the cosmic birefringence scenario we investigated in this paper by precisely measuring the low-$\ell$ E-mode power spectrum. An improvement in birefringence angle measurements obtained from radio galaxies will also become important in the near future to explore possible sources of cosmic birefringence \cite{Naokawa:2025:DEprofile}. Other methods include using galaxy shapes~\cite{Yin:2024:galaxy} and the polarized Sunyaev–Zel’dovich effect~\cite{Hotinli:2022:pkSZ,Lee:2022:pSZ-biref,Namikawa:2023:pSZ}. These observations will further test cosmic birefringence.

%////////////////////////////////////////%
% BACK MATTER 
%////////////////////////////////////////%

\begin{acknowledgments}
We thank Neal Jackson and Ian Browne for helpful discussions. 
TN acknowledges support from JSPS KAKENHI Grant No. JP25K00996, and No. JP26H00405, and from JST EXPERT-J, Japan Grant No. JPMJEX2508. The Kavli IPMU is supported by World Premier International Research Center Initiative (WPI Initiative), MEXT, Japan. FN acknowledges the Forefront Physics and Mathematics Program to Drive Transformation (FoPM), a World-leading Innovative Graduate Study (WINGS) Program, the University of Tokyo. FN is also supported by JSPS KAKENHI Grant Number 24KJ0668, 20H05850, 23K20035, The ANRI Fellowship, PID2023-146372OB-I00, funded by MICIU/AEI/10.13039/501100011033 and by ERDF, EU. This work uses resources of the National Energy Research Scientific Computing Center (NERSC). 
\end{acknowledgments}

\section*{Data availability}
The data that support the findings of this article will be publicly available. 

% References %
%\bibliographystyle{apsrev}
\bibliographystyle{mybst}
\bibliography{cite}

\end{document}